# DISCUSSION OF: TREELETS—AN ADAPTIVE MULTI-SCALE BASIS FOR SPARSE UNORDERED DATA

By Peter J. Bickel[1] and Ya'acov Ritov[2]

*University of California and The Hebrew University of Jerusalem*

We divide our comments on this very interesting paper into two parts following its own structure:

1. The use of treelets in connection with the correlation matrix of $\mathbf{X} = (X_1, \ldots, X_p)^\mathsf{T}$ for which we have $n$ i.i.d. copies, or as the authors refer to it, "unsupervised learning."
2. The use of treelets as a step in best fitting the linear regression of $X_1$ on $(X_2, \ldots, X_p)^\mathsf{T}$.

**1. Unsupervised learning.** The authors' emphasis is on the method as a useful way of representing data analogous to a wavelet representation where $\mathbf{X} = \mathbf{X}(t)$ with $t$ genuinely identified with a point on the line and observation at $p$ time points, but where the time points have been permuted.

As such, this can be viewed as a clustering method which, from their examples, gives very reasonable answers. However, to make more general theoretical statements and to permit comparison to other methods, they necessarily introduce the model

$$\mathbf{X} = \sum_{j=i}^{K} U_j v_j + \sigma Z_j, \tag{1}$$

where $\mathbf{U} = (U_1, \ldots, U_K)^\mathsf{T}$ is an unobservable vector, the $v_j$ are fixed unknown vectors, and $\mathbf{Z} \sim N_p(0, J_p)$, where $J_p$ is the identity, $N_p$ is the $p$ dimensional Gaussian distribution, and $\mathbf{U}, \mathbf{Z}$ are independent.

At this point, we are a bit troubled by the authors' analysis. We believe a key point, that is only stressed implicitly by the authors, is that the population tree structure, as defined, is only a function of the population covariance

Received November 2007; revised November 2007.
[1]Supported in part by NSF Grant DMS-06-05236.
[2]Supported in part by an ISF grant.

matrix. This is clear at Step 1, and follows since the Jacobi transformations depend only on the covariance and variances of the coordinates involved. This raises a problematic issue. If **U**, and hence **X**, has a Gaussian distribution, then the structure as postulated in (1) is not identifiable, as in known in factor analysis. Consider, for instance, Example 2. If we redefine $U_j^* = U_j$, $j = 1, 2$, $v_3^* = c_1 v_1 + c_2 v_2$, and $U_3^* = 0$, we are at the same covariance matrix as in (19) with only two nonoverlapping blocks.

The treelets transform evidently gives a decomposition attuned to the authors' beliefs of a block diagonal population structure with high intrablock correlation. But the theoretical burden of exhibiting classes of covariance matrices, other than ones whose eigenvectors are not only orthogonal but have disjoint support, and for which some version of sparse PCA cannot be utilized just as well, remains.

This is an insurmountable problem for any population parameter which is a function only of the covariance matrix.

A second difficulty, special to the treelets parameter $T(\Sigma)$, is that it is not defined uniquely for $\Sigma$ for which the maximal off diagonal correlation is not uniquely assumed. This is reflected in the authors' discussion in Section 3.1 of the possible instability of the empirical tree. In this context, we don't understand their statement that inferring $T(\Sigma)$ is not the goal. If not, what is?

This issue makes comparison to the other methods difficult. As they state any of the several methods for sparse PCA, for example, d'Aspremont et al. (2007), Johnstone and Lu (2008), would yield the same answer as theirs for their Example 1.

But is there a way of proceeding which teases out explicitly structures such as in (19) without limiting oneself to the covariance matrix? Suppose that we can write $\mathbf{U} = B\mathbf{e}$, where $\mathbf{e} = (e_1, \ldots, e_K)^\mathsf{T}$ is a vector of independent not necessarily identically distributed variables, such that *at most one of them is Gaussian*. That is, we assume the factor loading themselves are obtained structurally. Then we can write for $i = 1, \ldots, n$, $j = 1, \ldots, p$, $X_{ij} = \sum_{l=1}^{K} c_{jl} e_{il} + \sigma Z_{ij}$, where $C = [C_{jl}]$ is a $p \times K$ matrix, the $Z_{ij}$ are i.i.d. $N(0, 1)$, and $\mathbf{e}_i = (e_{i1}, \ldots, e_{iK})^\mathsf{T}$ are independent as above. Here, $C = VB$, where $V = (v_1, \ldots, v_k)$. We conjecture that if $p, n \to \infty$ with $K$ fixed, and the columns of $C$ are sparse, we can recover $C$ up to a scale multiple of each row, and a permutation of the columns. Work on this conjecture is in progress.

**2. Supervised learning.** Can we select variables based on the $X$, the predictor variables, themselves? The tempting answer is yes (e.g., using PCA). The theoretical answer is no ($Y$ can be a function of each component). The practical answer is at most a cautious yes; cf. Cook (2007) for a recent discussion. However, one should be careful to justify working with the predictions without the $Y$, since current regression methods permit one to handle models with almost exponentially many variables.

The LASSO type of estimator can handle sparse models. However, sparsity is an elusive property, since the LASSO can deal with sparsity in a given basis, while a sparse representation may exist only in some other basis. Treelets are proposed as a method which enriches the description of the model, and gives the user an over-rich collection of vectors which span the Euclidean space. Hopefully the tree cluster features are rich enough so the model can be approximated by the linear span of relatively few, say, no more than $o(n/\log n)$ terms.

The suggested algorithm deals with complexity by serial optimization in a fashion similar to standard model selection methods (e.g., forward selection), boosting, etc. It is not clear to us why the authors select the variables from one level and not from their union, since again modern methods can deal with any polynomial number of regressors.

To asses performance of the algorithm, we considered a simple version of the authors' supervised errors-in-variables model, but in an asymptotic setting. Suppose we observe $n$ i.i.d. replicates from the distribution of $(Y, X_1, \ldots, X_p)$, where $p = p_n$ and

$$Y = \gamma Z + \varepsilon,$$
$$X_i = c_p Z + \eta_i, \qquad i = 1, \ldots, p,$$

where $\varepsilon, Z \sim N(0,1)$, $\eta_i \sim N(0, \sigma_i^2)$, all independent. This is a classical error in variables model, where the $X_i$ are independent observations on $Z_i$ and the best predictor is given by

$$\hat{y}(X) = \frac{\gamma c_p}{1 + c_p^2 \sum_{i=1}^p \sigma_i^{-2}} \sum_{i=1}^p \sigma_i^{-2} X_i.$$

Consider first $c_p = p^{-1/2}$, with all $\sigma_i = 1$, $\gamma \neq 0$ and, in particular, $c_p^2 \times \sum_{i=1}^p \sigma_i^{-2} = 1$. In this case all variables are interesting, and have the same weight for prediction. However, the covariance matrix of $X$ has all diagonal terms greater than 1, and all off diagonal terms are $p^{-1}$. This model is not sparse—for instance, in the sense of El Karoui (2008), and is also inaccessible to regularized covariance estimation. The Treelet Algorithm will not be able to find this term. This model is significantly different from the null, and a consistent predictor exists given known parameter values. However, no standard general purpose algorithm will be able to deal with this model. A small set of simulations show that, in fact, there is a range of values of $c_p$ for which PCA works better than treelets. However, for larger values of $c_p$, treelets work surprisingly well.

The restriction to a basis of a relatively small collection of transform variables is a limitation. In Bickel, Ritov and Tsybakov (2008) a general methodology was suggested for construction of a rich collection of basis functions. Formally, we consider the following hierarchical model selection method. For a set of functions $\mathcal{F}$ with cardinality $|\mathcal{F}| \geq K$, let $\mathcal{MS}_K$ be some procedure to select $K$ functions out of $\mathcal{F}$. We denote by $\mathcal{MS}_K(\mathcal{F})$ the selected subset of $\mathcal{F}$, $|\mathcal{MS}_K(\mathcal{F})| = K$, $K = n^\gamma$ for some $\gamma < \infty$. Define $f \oplus g$ to be the operator combining two base variables, for instance, multiplication. The procedure is defined as follows:

(i) Set $\mathcal{F}_0 = \{X_1, \ldots, X_p\}$.
(ii) For $m = 1, 2, \ldots$, let

$$\mathcal{F}_m = \mathcal{F}_{m-1} \cup \{f \oplus g : f, g \in \mathcal{MS}_K(\mathcal{F}_{m-1})\}.$$

(iii) Continue until convergence is declared. The output of the algorithm is the set of functions $\mathcal{MS}_K(\mathcal{F}_m)$ for some $m$.

Bickel, Ritov and Tsybakov consider $f \oplus g = fg$, since they consider models with interaction. The treelets construction is similar to this one, with each step yielding two new functions, which result from PCA applied to a pair of variables. There is one essential difference between our approach and the treelets algorithm. We also keep at each step the complexity of the overdetermined collection in check, but let the complexity increase with the increase with levels.

DEPARTMENT OF STATISTICS  
UNIVERSITY OF CALIFORNIA  
BERKELEY, CALIFORNIA 94720-3860  
USA  
E-MAIL: bickel@stat.berkeley.edu

DEPARTMENT OF STATISTICS  
THE HEBREW UNIVERSITY OF JERUSALEM  
JERUSALEM  
ISRAEL  
E-MAIL: yaacov.ritov@gmail.com